\documentclass[aps,twocolumn,showpacs,preprintnumbers,superscriptaddress,nofootinbib]{revtex4}

\usepackage{subfigure,graphicx,epsfig,amsmath,amsfonts,amssymb,xcolor,slashed,ulem}

\newcommand{\be}{\begin{equation}}
\newcommand{\ee}{\end{equation}}
\newcommand{\ba}{\begin{eqnarray}}
\newcommand{\ea}{\end{eqnarray}}

\begin{document}

\title{ The \boldmath$\Lambda_b \to J/\psi ~ K ~ \Xi$ decay and the higher order chiral terms of the meson baryon interaction.}

\date{\today}

\author{A. Feijoo}
\email{feijoo@ecm.ub.edu}
\affiliation{Departament d'Estructura i Constituents de la Mat\`eria and Institut de Ci\`encies del Cosmos, Universitat de Barcelona, Mart\'i i Franqu\`es 1, 08028 Barcelona, Spain}

\author{V.K. Magas}
\affiliation{Departament d'Estructura i Constituents de la Mat\`eria and Institut de Ci\`encies del Cosmos, Universitat de Barcelona, Mart\'i i Franqu\`es 1, 08028 Barcelona, Spain}

\author{A. Ramos}
\affiliation{Departament d'Estructura i Constituents de la Mat\`eria and Institut de Ci\`encies del Cosmos, Universitat de Barcelona, Mart\'i i Franqu\`es 1, 08028 Barcelona, Spain}

\author{E.~Oset}
\affiliation{Departamento de
F\'{\i}sica Te\'orica and IFIC, Centro Mixto Universidad de Valencia-CSIC Institutos de Investigaci\'on de Paterna, Aptdo.22085, 46071 Valencia, Spain}


\begin{abstract} 

We study the weak decay of the $\Lambda_b$ into  $ J/\psi ~ K ~ \Xi$ and 
$J/\psi ~ \eta ~ \Lambda$ states,  and relate these processes to the
$\Lambda_b  \to J/\psi ~ \bar K ~N$ decay mode.
The elementary weak transition at the quark level proceeds via the creation of a $J/\psi$ meson and 
an excited $sud$ system with $I=0$, which upon hadronization leads to $\bar K N$ or $\eta \Lambda$ pairs. These states undergo final state interaction in coupled channels and produce a final meson-baryon pair. The $K \Xi$ state only occurs via rescattering, hence making the $\Lambda_b \to J/\psi ~ K ~ \Xi$ process very sensitive to the details of the meson-baryon interaction in strangeness $S=-1$ and isospin $I=0$. We show that the corresponding invariant mass distribution is dominated by the next-to-leading order terms of the chiral interaction. The $I=0$ selectivity of this decay, and its large sensitivity to the higher order terms, makes its measurement very useful and complementary to the  $K^- p \to K \Xi$ cross section data. The rates of the $\Lambda_b \to J/\psi ~ K ~ \Xi$ and $\Lambda_b \to J/\psi ~ \eta ~ \Lambda$ invariant mass distributions are sizable compared to those of the $\Lambda_b  \to J/\psi ~ \bar K ~N$ decay, which is measured experimentally, thus, we provide arguments for an experimental determination of these decay modes that will help us understand better the chiral dynamics at higher energies.
\end{abstract}


\maketitle

\section{Introduction}

The meson-baryon interaction in the strangeness $S=-1$ sector has been one of the favorite grounds to test nonperturbative chiral dynamics. The existence of the $\Lambda(1405)$ resonance below the $\bar K N$ threshold makes the use of nonperturbative unitary schemes 
mandatory to study the  $\bar K N$ interaction with its coupled channels. The combination of chiral dynamics and unitarity in coupled channels has proved very efficient in describing this interaction, even using the chiral Lagrangians to lowest order \cite{Ecker:1994gg,Bernard:1995dp}. The scheme combining these two essential dynamical aspects is usually referred to as the chiral unitary approach and has been widely used, providing good agreement with data \cite{Kaiser:1995eg,Kaiser:1996js,angels,ollerulf,Lutz:2001yb,bennhold,Hyodo:2002pk,cola,
GarciaRecio:2002td,GarciaRecio:2005hy}. One of the common findings of this approach is the existence of two poles for the $\Lambda(1405)$, one narrow around 1420 MeV, and another one, not so precise in the mass and wider, around 1380 MeV \cite{ollerulf,cola}. There are many reactions supporting this two pole structure and reviews on the situation can be seen in \cite{review,luis1,Magas:2005vu}. More recently, it has found additional support from the analysis of the $\pi \Sigma$ photoproduction data  \cite{Niiyama:2008rt,Moriya:2012zz,Moriya:2013hwg} in \cite{luis1,luis2,maimeissner}.

  The aim for precision and the need to extend the approach to higher energies has motivated the introduction of higher order terms of the chiral Lagrangians in the kernel, or potential, of the meson-baryon interaction. New fits to extended data, that include the valuable results of the $K^- p$ kaonic atom  \cite{Bazzi:2011zj}, have been performed with the aim of determining the parameters of the higher order chiral potential \cite{Borasoy:2005ie,Oller:2006jw,Borasoy:2006sr,hyodonew,maimeissner}. A recent study \cite{Feijoo:2015yja} has also incorporated the data of the $K^- p \to K^+ \Xi^-, K^0 \Xi^0$ reactions, since they do not proceed from the lowest-order chiral lagrangian and, hence, they are especially sensitive to the higher order terms. Most of the data employed in all fits are coming from antikaon proton scattering and  therefore contain contributions from both isospin $I=0$ and $I=1$ components, an exception being the $\pi^0 \Sigma^0$ production channel, which selects $I=0$. The chiral lagrangian models can also be tested against data from photoproduction \cite{Niiyama:2008rt,Moriya:2012zz,Moriya:2013hwg}, from
$p p \to K^+ \pi^0 \Sigma^0$   \cite{Zychor:2007gf} and from $K^- p  \to \pi^0 \pi^0 \Sigma^0$  \cite{Prakhov:2004an} reactions, but the energies where there is available information are essentially below the $\bar K N$ threshold where the higher order terms are relatively unimportant, or can be easily accommodated by changes in the subtraction constants in the regularization of the loop functions of the different channels. This is why data filtering $I=0$ at high energies would be most welcome as a complement of the information one can obtain from $K^- p$ scattering data. 
One such opportunity arises from the weak decay of the $\Lambda_b$ into states containing a $J/\Psi$ and meson-baryon pairs, measured by the CDF \cite{Aaltonen:2010pj} and LHCb \cite{Aaij:2013oha, Aaij:2014zoa, Aaij:2015tga} collaborations. In particular, the  $\Lambda_b\to J/\Psi~K^-~ p$ decay has been employed very recently in \cite{Aaij:2015tga} to claim the presence of an exotic pentaquark charmonium state in the $J/\Psi\, p$ channel.
A recent theoretical study of the $\Lambda_b \to J/\psi ~K^- p (\pi \Sigma)$ decay has been performed \cite{rocamai}, finding that this type of reaction does filter the final meson-baryon components in $I=0$.

In the present work we focus in the study of the $\Lambda_b \to J/\psi  ~K  ~\Xi$ and  
the $\Lambda_b \to J/\psi ~ \eta ~ \Lambda$ decay processes, since they are very sensitive to the details of the meson-baryon interaction at high energies, in particular to the higher order terms, and may help us improve our knowledge on the parameters of the chiral lagrangian.
The weak decay mechanism in these reactions is the same as that producing  $J/\psi$ and  $K^- p$ 
or $\pi \Sigma$, except that different channels are chosen in the final state interaction of the very few meson-baryon states which are allowed to be produced in a primary step by the selection rules.
More specifically, the $\Lambda_b$ decays weakly into $J/\psi$ and three quarks that hadronize to produce the primary meson-baryon components, which turn out to be $\bar K N $ and $\eta \Lambda$. The final state interaction of these states in coupled channels allows the production of $K \Xi$  in the case of the  $\Lambda_b \to J/\psi  ~ K  ~\Xi$ decay. Thus, not having the $K \Xi$ pair produced in the first step, it comes from rescattering of meson-baryon components and hence this decay process depends strongly on the behavior of the meson-baryon interaction. The coupled-channel models employed in the present study are based on the chiral lagrangian up to next-to-leading order with the parameters fixed in \cite{Feijoo:2015yja} to antikaon proton scattering data, including the $K\Xi$ production channels. When implemented in the decay processes studied in this work, we actually confirm their sensitivity to the meson-baryon lagrangian, in particular to its next-to-leading order terms, not only on the invariant mass decay rate of the $\Lambda_b \to J/\psi  ~K  ~\Xi$ process, but also on that of the $\Lambda_b \to J/\psi ~ \eta ~ \Lambda$ one, even if this process also receives contributions from the primary weak decay. 

The paper is organized as follows. In Sect.~\ref{sec:formalism} we present the formalism, describing the weak transition process and the implementation of final state interactions in Subsect.~\ref{subsec:weak}, and giving the details of the employed meson-baryon interaction in  Subsect.~\ref{subsec:strong}. Our results for the invariant mass distributions of the $\Lambda_b \to J/\psi  ~K ~  \Xi$ and  
the $\Lambda_b \to J/\psi ~ \eta ~ \Lambda$ processes are discussed in Sect.~\ref{sec:results}, where they are compared to those for the related  $\Lambda_b \to J/\psi ~ \pi~\Sigma, J/\psi ~ {\bar K}~N$ decays. Our concluding remarks are given in Sect.~\ref{sec:conclusions}.

\section{Formalism}\label{sec:formalism}

\subsection{The \boldmath$\Lambda_b\to J/\psi~M~B$ process}
\label{subsec:weak}  
\begin{figure*}[t]
\begin{minipage}{0.48\linewidth}
 \includegraphics[width=\linewidth]{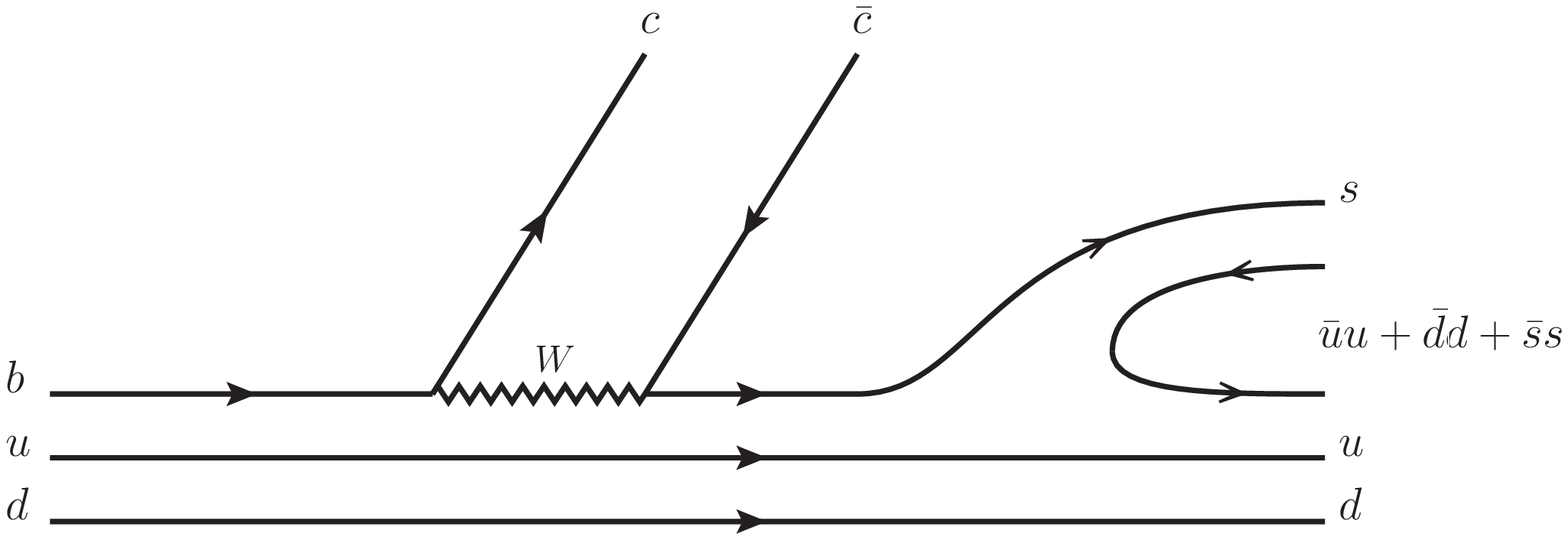}
$\hspace{0.05\linewidth}
\underbrace{\hspace{0.5\linewidth}}_{\text{Weak decay}}
\underbrace{\hspace{0.35\linewidth}}_{\text{Hadronization}}
\hspace{0.1\linewidth}$
\caption{Diagrams describing the production of a meson-baryon pair from the weak decay 
${\Lambda_b\to\Lambda\,J/\psi}$ 
through a hadronization mechanism. The full and serrated lines correspond to 
quarks and the $W$-boson, respectively.}\label{fig:weak}
\end{minipage}
~~~
\begin{minipage}{0.48\linewidth}
\vspace{+1cm}
\includegraphics[width=\linewidth]{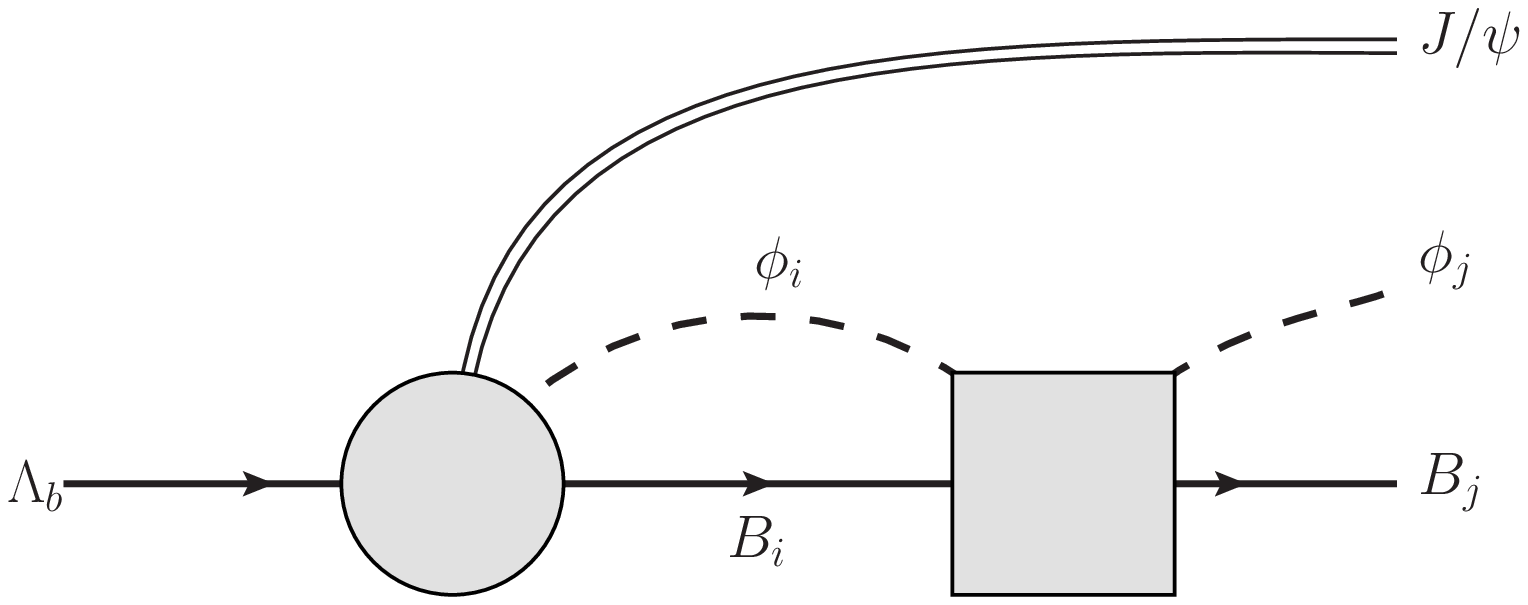}
\vspace{-0.1cm}
\caption{Final-state interaction of the meson-baryon pair, where the double, 
full and dashed lines denote 
the $J/\psi$, the baryons and the pseudoscalar mesons, respectively. 
The shaded circle and square stand for the production mechanism of the $J/\psi B_i\phi_i$, 
as depicted in Fig.~\ref{fig:weak}, and the meson-baryon scattering 
matrix $t_{ij}$, respectively}\label{fig:full}
\end{minipage}
\end{figure*}

In the decay of the $\Lambda_b$ into $J/\psi$ at the elementary quark level we must bear in mind that the $q\to q^\prime$ transitions at the $Wqq^\prime$ vertices are determined by the Kobayashi-Maskawa matrix elements \cite{Chau:1982da}. The $u\to d$ and $c \to s$ transitions are given by the cosine of the Cabibbo angle, $\cos \theta_C$, thus being Cabibbo favored, the $b\to c$ transition proceeds via $A \sin^2\theta_C$ and is Cabibbo suppressed, while the transition $b \to u$ would be doubly Cabibbo suppressed \cite{Wolfenstein:1983yz}. At the quark level, the Cabibbo favored mechanism for $J/\psi$ production is depicted by the first part of the diagram of Fig. \ref{fig:weak}. This corresponds to internal emission in the classification of topologies of \cite{Chau:1982da}, and is also the dominant mechanism in the related $\bar B^0 \to J/\psi~ \pi~ \pi$ decay \cite{stone,liang,melawei}. As we can see in the figure, a $sud$ state is obtained after the weak decay. The next step consists in the hadronization of this final three quark state by introducing a $\bar q q$ pair with the quantum numbers of the vacuum, $\bar u u+ \bar d d+ \bar s s$, which will then produce an initial meson-baryon pair. The final state interaction of this pair will produce a  final meson-baryon state in S-wave, which has $J^P=1/2^-$ quantum numbers. Since the hadronization is a strong interaction process, the $sud$  quark system produced in the weak process must have these quantum numbers.  Going back to the original $\Lambda_b$ state, one finds a $b$ quark and a $ud$ quark pair coupled to isospin $I=0$. The other observation from the decay mechanism is that the two $Wqq^\prime$ transitions occur in the same quark line involving the initial $b$ and the final $s$ quarks, thus, in this reaction the $u$ and $d$ quarks act as spectators. This means that the $ud$ pair in the final $sud$ state after the weak decay has $I=0$ and, since the $s$ quark also has $I=0$, the final three-quark system has total $I=0$. Hence, even if the weak interaction allows for isospin violation, in this case the process has filtered $I=0$ in the final state. Since isospin is conserved in the strong hadronization process and in the subsequent final state rescattering interaction, the final meson-baryon component also appears in $I=0$. 

  Another observation concerning the hadronization is that, since the $sud$ quark state after the weak decay has $J^P=1/2^-$ and the $ud$ quarks have the same quantum numbers as in the original $\Lambda_b$ state ($J^P=1/2^+$ each) in an independent quark model used  for the argumentation, it is the $s$ quark the one that must carry the minus parity, which would correspond to an $L=1$ orbit of a potential well. Since in the final meson-baryon states all the quarks will be in their ground state with $L=0$, the $s$ quark must be necessarily involved in the hadronization, as depicted in Fig. \ref{fig:weak}. Then it is easy to see that we would have a $s \bar u$ meson plus a $uud$ baryon, or $s \bar d$ and $dud$ or $s \bar s$ and $sud$. These configurations correspond to $K^- p$, $\bar K^0 n$ or $\eta (\eta') \Lambda$ ($\eta \Sigma$ cannot appear because it has $I=1$). The weights of these components remain to be determined, for what the quark structure for the baryons of the octet is needed. This  is simple and is shown in detail in \cite{rocamai}, with the result that the $sud$ state after the weak decay is given by 
\begin{equation}
|H\rangle=|K^-p\rangle+|\bar K^0 n\rangle
-\frac{\sqrt{2}}{3}|\eta\Lambda\rangle+\frac{2}{3}|\eta'\Lambda\rangle\,,
\label{eq:Hflav}
\end{equation}
or, equivalently, that the different possible meson-baryon pairs are created with a weight $h_i$, given by
\begin{align*}
&h_{\pi^0\Sigma^0}=h_{\pi^+\Sigma^-}=h_{\pi^-\Sigma^+}=0\,,~h_{\eta\Lambda}=
-\frac{\sqrt{2}}{3}\,,\\
&h_{K^-p}=h_{\bar K^0n}=1\,,~h_{K^+\Xi^-}=h_{K^0\Xi^0}=0\,.
\end{align*}

As is customary is these studies we neglect the $\eta' \Lambda$ component, and we only have primary $K^-p$, $K^0n$ or $\eta\Lambda$ production. We can see that a $K \Xi$ pair is not produced in the first step. 

Next, one must incorporate the final state interaction of these meson-baryon pairs, which is depicted in Fig. \ref{fig:full}. The matrix element for the production of the final state, $j$, is given by 
\begin{align}\label{eqn:fullamplitude}
\mathcal{M}_{j}(M_{\rm inv})=V_p\left( h_j+\sum_{i}h_iG_i(M_{\rm inv})\,t_{ij}(M_{\rm inv}) \right)\,,
\end{align}
where  $G_i$ denotes the one-meson-one-baryon loop function, chosen in 
accordance with the model for the scattering matrix $t_{ij}$ that will be described in the next section, and  $M_{\rm inv}$
is the invariant mass of the meson-baryon system in the final state. The factor $V_p$, which includes the common dynamics of the production of the different pairs, is unknown and we take it as constant.

Finally, the invariant mass distribution $\Lambda_b\to
J/\psi\,\phi_j\, B_j$ reads
\begin{align}\label{eqn:dGammadM}
\frac{d\Gamma_j}{dM_{\rm inv}}(M_{\rm inv})
=\frac{1}{(2\pi)^3}\frac{M_j}{M_{\Lambda_b}} p_{J/\psi} \, p_j\left|\mathcal{M}_{j}(M_{\rm inv})\right|^2\,,
\end{align}
where $p_{J/\psi}$ and $p_j$ stand for the modulus of the 
three-momentum
of the $J/\psi$ in the $\Lambda_b$ rest-frame and the modulus of the center-of-mass
three-momentum in the final meson-baryon system, respectively. The mass of the
final baryon is denoted by $M_j$.

\subsection{Summary of the Barcelona model for meson-baryon scattering}\label{sec:mmmodel}
\label{subsec:strong}
In this section we describe the model developed recently in Ref.~\cite{Feijoo:2015yja} with the aim of improving upon the knowledge of the chiral meson-baryon interation at next-to-leading (NLO) order in the strangeness $S=-1$ sector. The parameters of the Lagrangian were fitted to a large set of experimental scattering data in different two-body channels, as well as to branching ratios at threshold, and to the precise SIDDHARTA value of the energy shift and width of kaonic hidrogen \cite{Bazzi:2011zj}. Novel to other works in the literature, the model was also constrained to reproduce the $K^- p\to K^+\Xi^-, K^0\Xi^0$ reactions, since they  become  especially sensitive to the NLO terms, as they cannot proceed with the Lagrangian at lowest order, except indirectly via unitarization contributions.

The results presented in Ref.~\cite{Feijoo:2015yja} clearly established
the sensitivity of the NLO Lagrangian to
the  $K^- p \to K\Xi$ reactions, thus yielding better constrained parameters. That work also
investigated the influence of high spin hyperon resonances to the $K^- p \to K\Xi$ amplitudes.
The resonant terms helped in improving the description of the scattering data, including also the differential cross sections of the $K\Xi$ production reactions. In addition, by absorbing certain structures of the cross section, the inclusion of resonant contributions helped in obtaining more precise values of the low energy constants of the chiral unitary model.

More especifically, and similarly to many other chiral unitary models, the meson-baryon amplitudes built up in Ref.~\cite{Feijoo:2015yja} start from a kernel obtained from the SU(3) chiral Lagrangian up to NLO:
\begin{equation}
v_{ij}=v^{\scriptscriptstyle WT}_{ij}+v^{\scriptscriptstyle NLO}_{ij} 
\label{eq:kernel}
\end{equation}
where
\begin{equation}
v^{\scriptscriptstyle WT}_{ij}=
 - \frac{C_{i j}(2\sqrt{s} - M_{i}-M_{j})}{4 f^2}\! N_{i} N_{j}
\end{equation}
and
\begin{equation}
v^{\scriptscriptstyle NLO}_{ij} =\frac{D_{ij}-2(k_\mu k^{\prime\,\mu})L_{ij}}{f^2}\! N_{i} N_{j}\ ,
\end{equation}
with
\begin{equation}
N_{i}=\sqrt{\frac{M_i+E_i}{2M_i}},\,\, N_{j}=\sqrt{\frac{M_j+E_j}{2M_j}} \nonumber \ .
\end{equation}\
The indices $i,j$ stand for any of the ten meson-baryon channels in the neutral $S=-1$ sector: $K^-p$, $\bar{K}^0 n$, $\pi^0\Lambda$, $\pi^0\Sigma^0$, $\pi^-\Sigma^+$, $\pi^+\Sigma^-$, $\eta\Lambda$, $\eta\Sigma^0$, $K^+\Xi^-$ and $K^0\Xi^0$. The quantities $M_i,M_j$ and $E_i,E_j$ are the masses and energies, respectively, of the baryons involved in the transition. The lowest order kernel $v^{\scriptscriptstyle WT}_{ij}$, or Weinberg-Tomozawa (WT) term, is given in terms of the pion decay constant $f$ and a matrix of coefficients $C_{ij}$. The other low energy constants are embedded in the matrices  
$D_{ij}$ and $L_{ij}$ of the NLO term, $v^{\scriptscriptstyle NLO}_{ij}$. These matrices are well known and can be found, for example, in the appendices of Ref.~\cite{Feijoo:2015yja}.  It is important to stress that the inclusion of NLO terms in the Lagrangian implies an increase, from one to eight, of the chiral interaction parameters to be fitted, the uncertainty of which was limiting the predictive power of the NLO models. An important step to remedy this situation was done in the study of Ref.~\cite{Feijoo:2015yja} by employing the data on $K \Xi$ production reactions, which are especially sensitive to the NLO contributions.

The interaction kernel cannot be employed perturbatively to describe
the scattering of $\bar{K}N$ states, since they couple strongly to many other channel states generating the two poles of the $\Lambda(1405)$ and the $\Lambda(1670)$ in $I=0$, plus other resonances in other sectors. A nonperturbative resummation is needed to describe this system and the present model employs the Bethe-Salpeter equation in its on-shell factorized form:
\begin{equation}
t_{ij} =v_{ij}+v_{il} G_l t_{lj}  \ .
\label{LS}
\end{equation} 
The loop function $G$ stands for a diagonal matrix with elements: 
\begin{equation} \label{Loop_integral}
G_l={\rm i}\int \frac{d^4q_l}{{(2\pi)}^4}\frac{2M_l}{{(P-q_l)}^2-M_l^2+{\rm i}\epsilon}\frac{1}{q_l^2-m_l^2+{\rm i}\epsilon} \ ,
\end{equation} 
where $M_l$ and $m_l$ are the baryon and meson masses of the $``l"$ channel. The logarithmic divergences of the loop function are treated within dimensional regularization:
\begin{eqnarray}
& G_l = &\frac{2M_l}{(4\pi)^2} \Bigg \lbrace a_l+\ln\frac{M_l^2}{\mu^2}+\frac{m_l^2-M_l^2+s}{2s}\ln\frac{m_l^2}{M_l^2} + \nonumber \\ 
 &     &\frac{q_{\rm cm}}{\sqrt{s}}\ln\left[\frac{(s+2\sqrt{s}q_{\rm cm})^2-(M_l^2-m_l^2)^2}{(s-2\sqrt{s}q_{\rm cm})^2-(M_l^2-m_l^2)^2}\right]\Bigg \rbrace ,  
 \label{dim_reg}    
\end{eqnarray}
where we have taken $\mu=1$ GeV as regularization scale, and $a_l$ are subtraction constants, which are also taken as free parameters respecting isospin symmetry. We note that this model neglects the $s$ and $u$-channel diagrams involving the coupling of the meson-baryon channel to an intermediate ground state baryon, which have been shown to contribute very moderately \cite{ollerulf,Borasoy:2006sr,hyodonew}.

The chiral model was complemented with the explicit inclusion, in the $K^-p \to K^0\Xi^0,K^+ \Xi^-$ amplitudes, of two high spin resonances.  From the possible candidates listed in the PDG \cite{PDG}, and according to other resonance-based models \cite{Sharov:2011xq,jackson},
the $\Sigma(2030)$ and  the $\Sigma(2250)$ resonances were selected. The spin and parity $J^\pi =7/2^+$ of the $\Sigma(2030)$ are well established. Those of the $\Sigma(2250)$ are not known,  but the choice
$J^\pi =5/2^-$ was adopted out of the two most probable assignments, $5/2^-$ or $9/2^-$  \cite{PDG,Sharov:2011xq}. With the resonances included, the amplitudes connecting $K^- p$, $\bar{K}^0 n$ states with $K^+ \Xi^-$, $K^0 \Xi^0$ ones should be replaced as
\begin{equation}
t_{ij} \to  t_{ij}
+\frac{1}{\sqrt{4M_pM_\Xi} }\, t_{ij}^{{5/2}^-}\, +\frac{1}{\sqrt{4M_pM_\Xi} }\, t_{ij}^{{7/2}^+}\ .
\label{T_reso}
\end{equation}
The resonant amplitudes $ t_{ij}^{R}$ contain the appropiate Clebsh-Gordan coefficients projecting the $i$ and $j$ states into the isospin value $I=1$ of the $5/2^-$ and $7/2^+$ resonances included. The attempt in \cite{Feijoo:2015yja} of incorporating the additional effect of the  $I=0$ $J^\pi=3/2^-$ $\Lambda(1890)$ resonance did not produce a substantial improvement in the description of the data. More details on the implementation of the resonant terms can be found in Ref.~\cite{Feijoo:2015yja}.

Once the amplitudes $t_{ij}(M_{\rm inv})$ are known, we can insert them in Eq.~(\ref{eqn:fullamplitude}) to determine the matrix element for the decay of the $\Lambda_b$ into $J/\Psi$ and a particular final meson-baryon state $j$.

The results presented in this work will be based on two of the models developed in Ref.~\cite{Feijoo:2015yja}. The first one employs only the dynamics of the chiral Lagrangian up to NLO and will be referred to as Model 1. The second one, denoted by Model 2, includes the additional contribution of the two resonances. The low energy parameters, subtracting constants and, in the case of Model 2, the couplings, masses, widths and form-factor cut-offs of the resonances, were fitted to reproduce $K^- p $ threshold branching ratios, as well as the $K^- p $ scattering data for elastic an inelastic processes, including the total and differential cross sections of the $K^-p \to K^0\Xi^0, K^+ \Xi^-$ reactions. The details on how the observables are reproduced can be seen in Ref.~\cite{Feijoo:2015yja}. One can also find in Table VI of that work the resulting values of the parameters, noting that Models 1 and 2 are named there as NLO* and NLO+RES, respectively. Here we only give explicitly the value of the $\chi^2_{\rm d.o.f.}$, 1.48 for Model 1 and 1.05 for Model 2, which informs in a global way on the goodness of these fits. We note that fit denoted by WT+RES  in \cite{Feijoo:2015yja}, which did not incorporate the NLO terms of the chiral Lagrangian, was of much lower quality, producing a $\chi^2_{\rm d.o.f.}$ of 2.26.

\section{Results}\label{sec:results}

\begin{figure*}[!htb]
\centering
  \includegraphics[width=0.9\textwidth]{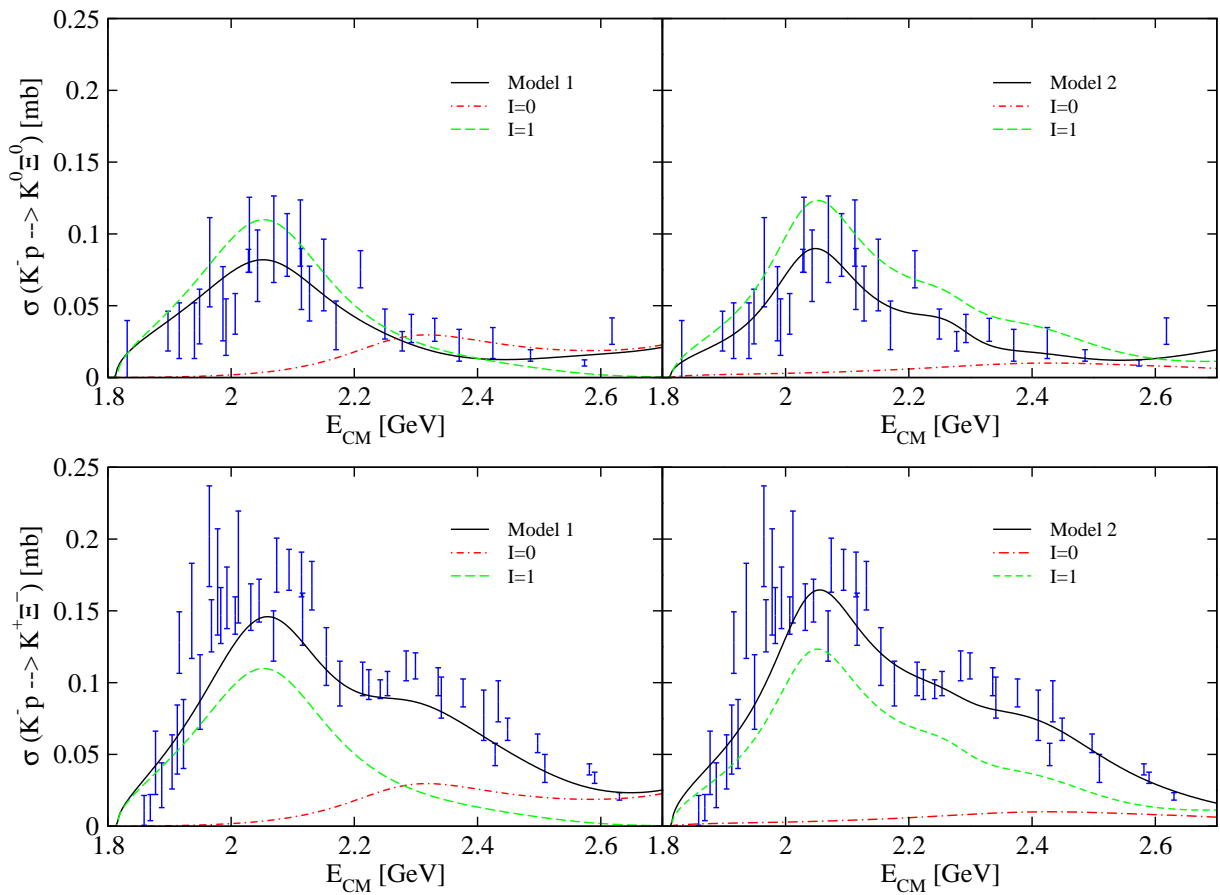}
\caption{(Color online) The total cross sections of the $K^- p\to K^0 \Xi^0$ reaction (top row) and the $K^- p\to K^- \Xi^+$ reaction (bottom row) for the two different models (Model 1 and Model 2) discussed in the text. The solid lines show the results of the full amplitude, while the dashed  and dash-dotted lines denote the $I=1$ and $I=0$ contributions, respectively. Experimental data are from \cite{exp_1, exp_2, exp_3, exp_4, exp_5, exp_6, exp_7}.} 
  \label{fig:iso}
\end{figure*}
We start this section by presenting in Fig.~\ref{fig:iso} the cross section data of the  $K^- p\to K^0 \Xi^0$ reaction (top panels) and of the $K^- p\to K^- \Xi^+$ reaction (bottom panels), obtained employing Model 1 (left panels) or Model 2 (right panels). The figure shows the complete results (solid lines), as well as the results where only the isospin $I=1$ component (dashed lines) or the $I=0$ one (dash-dotted lines) have been retained. It is interesting to see that, in both models, the $I=1$ component is dominant and is concentrated at lower energies. The smaller $I=0$ component at higher energies adds up destructively to the cross section in the case of the $K^- p\to K^0 \Xi^0$ reaction, while it contributes to enhance the cross section in the $K^- p\to K^- \Xi^+$ process. We note that the tree-level chiral contributions to these reactions come entirely from the NLO Lagrangian and, upon inspecting the size of the coefficients, their strength in the $I=0$ channel would be nine times larger than that in the $I=1$ channel. The reversed trend observed in Fig.~\ref{fig:iso} is a consequence of the unitarization in coupled channels with coupling coefficients determined by the fit and, consequently, by the data. 

As we can see in Fig.~\ref{fig:iso}, the contribution of $I=0$ in the $K^-p \to K\Xi$ cross section has a maximum around 2300~MeV for Model 1 or around 2400~MeV and less pronounced for Model 2, far from the peak of the data and of the complete amplitude, around 2050~MeV. The $K^- p\to K \Xi$ reactions contain a mixture of both isospin components, while the decay process $\Lambda_b \to J/\psi ~ K ~ \Xi$, studied in this paper, filters $I=0$ and therefore provides additional information to the one obtained from the scattering data.

\begin{figure}[!htb]
\centering
  \includegraphics[width=\linewidth]{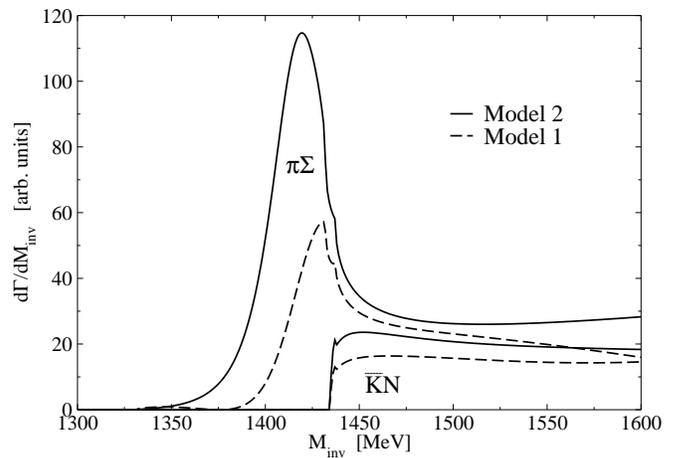}
\caption{Invariant mass distributions of $\pi\Sigma$ and $\bar{K} N$ states in the decay modes $\Lambda_b \to J/\psi ~ \pi ~\Sigma$ and  $\Lambda_b \to J/\psi ~ \bar{K} ~N$, for the two models discussed in the text: Model 1 (dashed lines) and Model 2 (solid lines). The units in the $y$ axis are obtained taking $V_p=1$. }
  \label{fig:KbarN}
\end{figure}
Since the models of \cite{Feijoo:2015yja} make a fitting to all $K^-p \to X$ data in a range from threshold to $K\Xi$ production, we start presenting, in Fig.~\ref{fig:KbarN},  what are the predictions of Model 1 and Model 2 for the decay reactions  $\Lambda_b \to J/\psi ~\bar{K} ~ N$ and $\Lambda_b \to J/\psi ~\pi ~ \Sigma$, already studied in \cite{rocamai}.  These are averaged distributions over the possible different charged states. 
We can see that the results of both models are similar to those found in \cite{rocamai}, with the shape of the $\pi \Sigma$ and $\bar K N$ distributions lying somewhat in between those of the Bonn and Murcia-Valencia models studied there (a different normalization is used in that work). We note that our $\pi \Sigma$ distributions shown in Fig.~\ref{fig:KbarN} stay over the ${\bar K}N$ ones, in contrast to what one observes in the models discussed in \cite{rocamai}, where the $\pi\Sigma$ distributions cross below the respective ${\bar K}N$ ones just above the threshold for ${\bar K}N$ states.  We have checked that this is a peculiarity of our NLO contributions, since we also obtain a crossing behavior when our interaction models are restricted to only the lowest order terms.  It is also interesting to see that the numerical results in Fig.~\ref{fig:KbarN} depend on the model, indicating their sensitivity on different parametrizations that fit equally well the  $K^-p \to X$ data. Actually, as seen from Eq.~(\ref{eqn:fullamplitude}), the rescattering term of the invariant mass distribution, which is dominant around the energy region of the $\Lambda(1405)$ resonance, depends not only on the strong scattering amplitudes, $t_{ij}$, but also on the loop functions, $G_i$. Since the fitting procedures consider the parameters of the meson-baryon interaction simultaneously with those of the loop functions, there is some freedom on the values of these loop functions obtained by different strong interaction models that produce equivalent scattering amplitudes. In addition, since the global parameter $V_p$ is unknown by us, the relevant information from this figure is the ratio of the $\pi \Sigma$ to $\bar{K}N$  distributions, at their respective maximum values for instance. We obtain ratios of 4.9 and 3.5 for Models 1 and 2, respectively. These values are of the order of those found
 in \cite{rocamai}. 

\begin{figure}[!htb]
\centering
  \includegraphics[width=\linewidth]{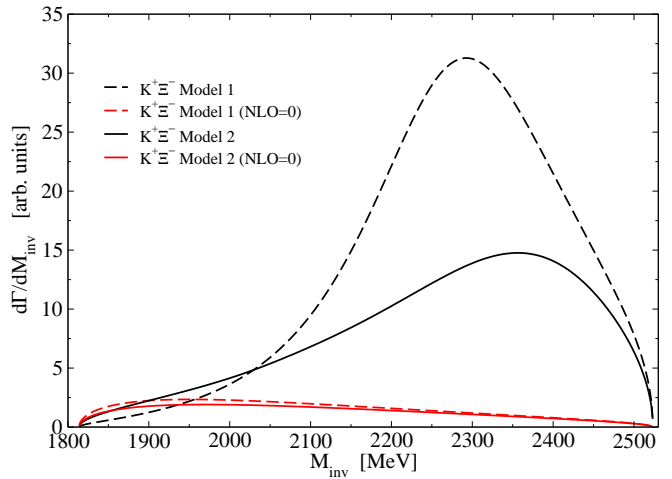}
\caption{(Color online) Invariant mass distributions of $K^+\Xi^-$ states produced in the decay $\Lambda_b \to J/\psi ~ K^+~\Xi^-$, obtained for the two models discussed in the text: Model 1 (dashed lines) and Model 2 (solid lines). The thin lower lines correspond to omitting the NLO terms of the potential. The normalization is the same as in Fig.~\ref{fig:KbarN}.}
  \label{fig:KXi}
\end{figure}

In Fig.~\ref{fig:KXi} we present the invariant mass distributions of the $K^+\Xi^-$ states from the decay process $\Lambda_b \to J/\psi ~ K^+ ~\Xi^-$. We do not show the distribution for the decay process $\Lambda_b \to J/\psi ~ K^0~ \Xi^0$, because, except for minor differences associated to the slightly different physical masses of the particles, it is identical to that of the charged channel, since these processes involve only the $I=0$ part of the strong meson-baryon amplitude. The fact that this decay filters the $I=0$ components makes the differences between Model 1 (thick dashed line) and Model 2 (thick solid line) to be more evident, not only in the strength but also in the shape of the invariant mass distribution. If, in order to eliminate the dependence on undetermined loop functions and on the unknown weak parameter $V_p$, we represented each $\Lambda_b \to J/\psi ~ K^+ ~\Xi^-$ distribution relative to its corresponding $\Lambda_b \to J/\psi ~ {\bar K} ~ N$ one shown in Fig.~\ref{fig:KbarN}, the difference would even be somewhat enhanced.
 Therefore, measuring the decay of the $\Lambda_b$ into $J/\psi ~ K^+~\Xi^-$ and into  J$/\psi ~ {\bar K} ~ N$ could help us discriminate between models that give a similar account of the scattering $K^- p \to K^0 \Xi^0, K^+ \Xi^-$ processes.  The figure also shows that the $I=0$ structure observed around 2300~MeV results from the terms of the NLO lagrangian. When they are set to zero, the invariant mass distributions of the two models, shown by the thin dashed and thin solid lines in Fig.~\ref{fig:KXi}, become small and structureless.

\begin{figure}[!htb]
\centering
  \includegraphics[width=\linewidth]{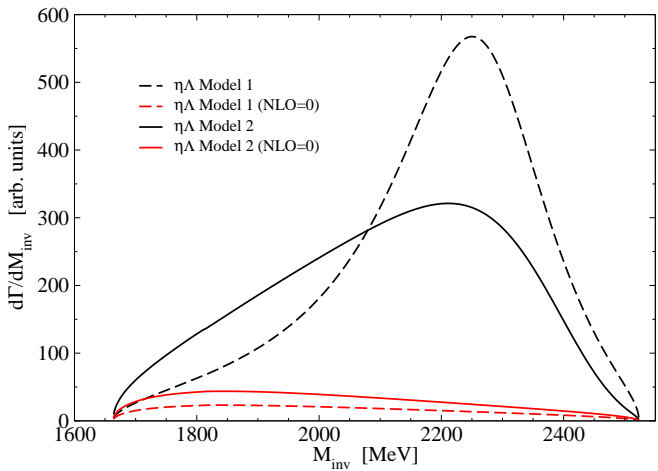}
\caption{(Color online) Invariant mass distributions of $\eta \Lambda$ states produced in the decay $\Lambda_b \to J/\psi ~ \eta ~\Lambda$, obtained for the two models discussed in the text: Model 1 (dashed lines) and Model 2 (solid lines). The thin lower lines correspond to omitting the NLO terms of the potential. The normalization is the same as in Fig.~\ref{fig:KbarN}.}
  \label{fig:etaL}
\end{figure}

We have observed a similar behaviour in the mass distributions of the reaction  $\Lambda_b \to J/\psi ~\eta ~\Lambda$ which are shown in Fig.~\ref{fig:etaL}. In this case, as the coefficient $h_{\eta\Lambda}$ does not vanish, we see from Eq.~(\ref{eqn:fullamplitude}) that the tree level term also contributes here, unlike the case of $K\Xi$ production. This makes the magnitude of the $\Lambda_b \to J/\psi ~\eta ~\Lambda$ mass distribution about twenty times bigger than that of the $\Lambda_b \to J/\psi ~ K ~ \Xi$ one.

\begin{figure}[!htb]
\centering
  \includegraphics[width=\linewidth]{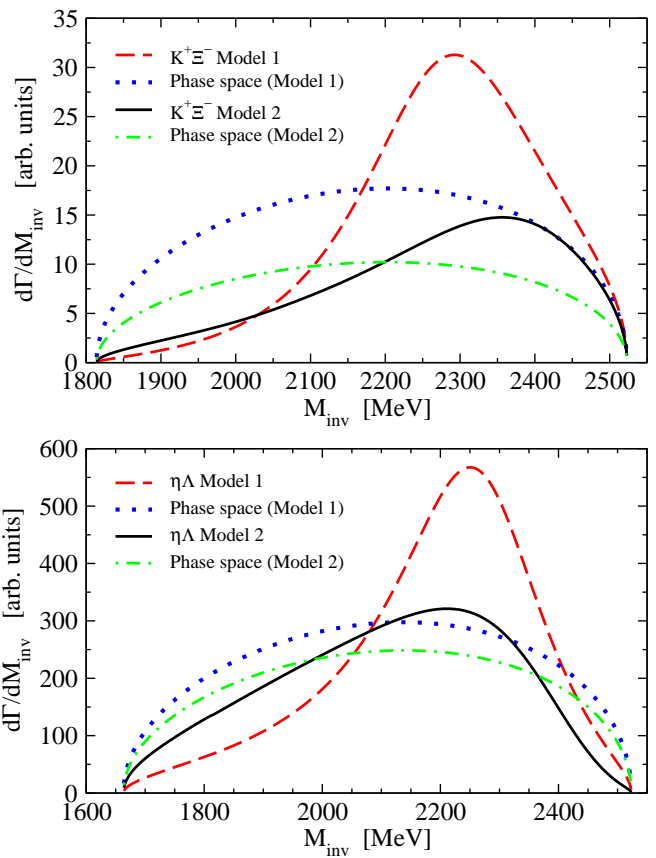}
\caption{(Color online) Comparison of the invariant mass distributions of $K^+\Xi^-$ states (upper panel) and $\eta \Lambda$ states (lower panel) states obtained with Model 1 (dashed lines) and Model 2 (solid lines) with a pure phase-space distribution (dotted lines)}
  \label{fig:PS}
\end{figure}

The invariant mass distributions from the $\Lambda_b \to J/\psi ~ K^+~\Xi^-$ and $\Lambda_b \to J/\psi ~ \eta~\Lambda$ decays obtained in Models 1 and 2 are compared with phase space in Fig.~\ref{fig:PS}. The phase-space distributions (dotted lines for  Model 1 and dash-dotted lines for Model 2) are obtained by taking the amplitude ${\cal M}_j$ as constant in Eq.~(\ref{eqn:dGammadM}) and normalizing to the area of the invariant mass distribution of the corresponding model. The comparison allows one to see that there are dynamical features in the meson-baryon amplitudes leading to a distinct shape of the mass distributions. In the case of Model 1, we observe a peak between  2250~MeV and 2300~MeV for both $\Lambda_b \to J/\psi ~ K^+~\Xi^-$ and $\Lambda_b \to J/\psi ~ \eta~\Lambda$ distributions. The peak resembles a resonance, but we should take into account that the limitation of the phase space at about 2500~MeV produces a narrower structure than that of the cross sections of the $K^-p \to K \Xi$ reactions, as we can see from the $I=0$ contribution in Fig.~\ref{fig:iso} (left panels), which is much broader.  Actually, the $I=0$ contribution of Model 2 to the cross sections of Fig.~\ref{fig:iso} (right panels)  does not indicate any particular structure, and the very different shapes that this model predicts for  $\Lambda_b \to J/\psi ~ K^+~\Xi^-$ and $\Lambda_b \to J/\psi ~ \eta~ \Lambda$ (see the thick solid lines in Fig.~\ref{fig:PS}), peaking at about 2400~MeV and 2200~MeV respectively, do not indicate the presence of a resonance since it would necessarily appear in both final states at the same energy. In order to find extra information concerning this issue, we have performed an extrapolation of the models to the complex plane, keeping the kernel potential $v_{ij}$ real, i. e. calculated at $Re(\sqrt{s})$,  but allowing the loop  functions $G_l$ to become complex. Employing this prescription, we do not find poles of the meson-baryon scattering amplitude in the second Riemann sheet around these energies. 
In our models, it is the energy dependence in the parametrization of the next-to-leading order contribution and the interference of terms what creates this shape. In any case, what is clear is that the experimental implementation of this reaction will provide valuable information concerning the meson-baryon interaction at higher energies, beyond what present data of scattering has offered us. 

Although we have given the invariant mass distributions in arbitrary units, one should bear in mind that all the figures, from Fig.~\ref{fig:KbarN} to Fig.~\ref{fig:etaL} have the same normalization. Since measurements for the $\Lambda_b \to J/\psi ~ K^- ~ p$ reaction are already available from the CDF \cite{Aaltonen:2010pj} and LHCb \cite{Aaij:2013oha, Aaij:2014zoa,Aaij:2015tga} collaborations, the measurements of the reactions proposed here could be referred to those of the $\Lambda_b \to J/\psi ~ K^-~ p$ reaction and this would allow a direct comparison with our predictions. In this spirit, we note that the recent resonance analysis of \cite{Aaij:2015tga} shows a $\Lambda(1405)$ contribution which lies in between the distribution found by the Bonn model in \cite{rocamai} and that of the Murcia-Valencia model in \cite{rocamai} or the Barcelona models presented here.

\section{Conclusions}\label{sec:conclusions}

We have shown that the $\Lambda_b \to J/\psi ~\eta~ \Lambda$ and particularly the $\Lambda_b \to J/\psi~  K ~\Xi$ reactions provide very valuable information concerning the meson-baryon interaction in the $S=-1$ and isospin $I=0$ sector.
The dynamics of the reaction, where the light quarks of the $\Lambda_b$ play a spectator role, is such that it filters $I=0$ in the final state. This is so because the $u d$ quarks in the $\Lambda_b$ baryon necessarily couple to $I=0$ and the weak decay favors the $b \to c\bar{c}s$ transition, so there is an $s$-quark at the end of the weak process, which together with the $u d$ pair in $I=0$ gives a total isospin $I=0$.
Thus, these decays may offer complementary information to that obtained from $K^-p \to K \Xi$ scattering data, where both  $I=0$ and $I=1$ contributions combine to give the final results.

Our study is based on models of 
 $K^-p$ scattering that include the next-to-leading order terms of the chiral Lagrangian and some explicit $I=1$ resonances, 
which do not contribute directly to the studied decays but their inclusion does modify the parameters of Model 2 with respect to those of Model 1. Both models produce quite different invariant mass distributions for the decay of the $\Lambda_b$ into $K\Xi$ and $\eta \Lambda$ states, which are in turn quite different also from phase space, indicating the sensitivity of these processes to the strong internal dynamics. The differences from phase space are more visible in Model 1 for both decay processes and in Model 2 for the $\Lambda_b \to J/\psi ~ K ~ \Xi$ one.
The reason stems from the fact that the decay into $\eta \Lambda$ can proceed at tree level, while the selectivity of the $\Lambda_b$ decay processes producing the $J/\psi$  does not allow the formation of a $K\Xi$ pair in a primary step. This is only produced through rescattering of the $\bar{K}N$ and $\eta \Lambda$ primary components. Thus the $\Lambda_b \to J/\psi~ K ~\Xi$ reaction is directly proportional to the meson-baryon scattering amplitude, concretely to the $\eta \Lambda \to K \Xi$ and $\bar{K}N \to K \Xi$ components in $I=0$.
The theoretical models fitted to the $K^-p$ scattering data lead to  $\Lambda_b \to J/\psi~ K ~ \Xi$ mass distributions with a peaked structure around 2300~MeV, lying far away from the
one around 2050~MeV in the $K^-p \to K \Xi$ cross section, which is dominated by $I=1$. These models also predict sizable differences for the $\Lambda_b$ decay in the energy region of $K \Xi$ and $\eta \Lambda$ production, reflecting that the $I=0$ component of the meson-baryon interaction, which is the one playing a role in the $\Lambda_b$ decay processes studied here, is not very well constrained by the fitting to $K^- p \to K \Xi$ data.

All the features observed in the present work indicate that the actual measure of these observables would provide valuable information, novel so far, that would enrich our knowledge of the meson-baryon interaction and help us make progress in our understanding of hadron dynamics.

\section*{Acknowledgments}
We would like to acknowledge enlightening discussions with L. Roca. This work is partly supported by the Spanish Ministerio de Economia y Competitividad and
European FEDER funds under the contracts FIS2011-28853-C02-01 and FIS2011-24154, 
by the Generalitat Valenciana in the program Prometeo II-2014/068,
by the Ge\-ne\-ra\-li\-tat de Catalunya contract 2014SGR-401,
and 
by the Spanish Excellence Network on Hadronic Physics FIS2014-57026-REDT.

\end{document}